\documentclass{emulateapj}
\input epsf

\def\mdot{\dot{\rm M}}

\slugcomment{Accepted for publication in the {\em{Spitzer}} Edition of the Astrophyical Journal Supplement Series}

\shorttitle{IR Wind Spectrum of EZ CMa}
\shortauthors{Morris, Crowther, \& Houck}

\begin{document}


\title{{\em{Spitzer}}-IRS Spectroscopy of the Prototype Wolf-Rayet Star
EZ CMa (HD\,50896)}


\author{Patrick W. Morris\altaffilmark{1}}
\affil{{\em{Spitzer}} Science Center, IPAC, California Institute of Technology,
M/S 220-6, 1200 E.\ California Blvd., Pasadena CA 91125 \\
\tt{pmorris@ipac.caltech.edu} }

\author{Paul A. Crowther}
\affil{Department of Physics \& Astronomy,  University of 
Sheffield, Hicks Building, Hounsfield Rd,  Sheffield S3 7RH, United Kingdom} 

\and
\author{Jim R. Houck}
\affil{Astronomy Department, Cornell University, 106 Space 
Sciences Bldg., Ithaca, NY 14853}

\altaffiltext{1}{NASA {\em{Herschel}} Science Center, IPAC, Caltech, M/S 
100-22, Pasadena, CA 91125}

\setcounter{footnote}{1}






\begin{abstract}
We present mid-infrared {\it Spitzer}-IRS\footnote{The {\em{Spitzer}} Space Telescope 
is operated by the Jet Propulsion Laboratory, California Institute of Technology 
under NASA contract 1407. Support for this work was partly provided by NASA through an award 
issued by JPL/Caltech. PAC acknowledges financial support from the Royal Society.  The 
IRS was a collaborative venture between Cornell University and 
Ball Aerospace Corporation funded by NASA through the Jet Propulsion 
Laboratory and the Ames Research Center.} spectroscopy 
of the prototype WN star EZ CMa (HD\,50896, WN4b). 
Numerous stellar wind lines of He\,{\sc ii} are revealed,
plus fine-structure lines of [Ne\,{\sc iii}] 15.5$\mu$m and [O\,{\sc iv}]
25.9$\mu$m. We carry out a spectroscopic analysis of HD\,50896 allowing
for line blanketing and clumping, 
which is compared to the mid-IR observations. We make use of these stellar
properties to accurately derive Ne/He=1.2--1.8$\times 10^{-4}$ and 
O/He=4.2--4.8$\times 10^{-5}$ by number,
for the first time in an early WN star. In addition,
we obtain N/C$\sim$40 and N/O$\sim$50 by number, values 
in perfect agreement with current predictions 
for rotating massive stars at the end of interior hydrogen burning. 
\end{abstract}


\keywords{techniques: spectroscopic --- stars: individual (HD 50896) --- stars: atmospheres --- infrared: stars}



\section{Introduction}

Wolf-Rayet (WR) stars are descended
from the most massive O stars ($M_{\rm{init}} > 25 M_\odot$), and shed
core-processed material in thick, line-driven winds at such prodigious rates 
(typically 10$^{-5} - 10^{-4} M_\odot$ yr$^{-1}$) as to 
affect the evolution of the core and thus the lifetime of the star, prior
to ending in a supernova explosion (Meynet \& Maeder 2003). The prototype 
Wolf-Rayet star EZ CMa 
(HD~50896 = WR~6 in the catalogue of van der Hucht 2001) has been studied 
extensively at nearly all wavelengths accessible with ground-based 
facilities and space observatories, to understand the extreme stellar 
wind properties of these rare massive stars, and how they
interact with their local environment. 

At infrared wavelengths, 
the continuum opacity is principally determined by free-free 
scattering of electrons from helium ions, and the emergent radiation 
originates in the outer layers of the wind where the asymptotic velocity
is reached (e.g., Hillier et al. 1983).  These layers are also where critical 
electron densities and temperatures are reached to form fine structure lines 
of Ne, O, S, and other elements, that are instrumental in testing evolutionary 
model predictions of the surface abundances during different core burning 
stages. 

Elemental abundances obtained from the fine structure
lines depend on the stellar parameters, particularly the mass-loss 
rat, which is sensitive to the degree of clumping in the wind. 
The role played by mid-infrared spectra, in conjunction with
UV, optical, and near-IR spectroscopy, has been demonstrated for a few 
WN and WC stars observed with the SWS (Morris et al. 2000; 
Dessart et al. 2000).  Only the brightest WR stars could be observed with
the SWS, for sensitivity reasons, and observations of only
five WR stars produced spectra of suitable quality for modelling
beyond 4.5~$\mu$m.

In this Letter we present {\em{Spitzer}}-IRS spectra of 
the prototype early WN star 
EZ CMa. We carry out an analysis of the ultraviolet to mid-IR
spectroscopy, revealing stellar and wind parameters, with which
 Ne/He and O/He  abundance ratios are determined. Comparisons with 
theoretical expectations are presented. 
    
\section{Observations}\label{obs}


EZ CMa was observed with the {\em{Spitzer}} InfraRed Spectrometer (described
by Houck et al. 2004) on 11 November 2003, during the Science Verification phase, 
using all four of the IRS modules.  The data were acquired with the 
Spectral Mapping AOT, with the star observed at 3 to 5 discrete positions 
along the lengths (cross-dispersed axes) of the slits.  The $10 - 37.5~\mu$m 
data presented in this paper are from the Short and Long High (SH and LH) 
resolution modules ($R \equiv \lambda / \Delta\lambda \simeq 700$), and
$5.3 - 10~\mu$m data are from the Short Low (SL) module ($R \sim 70 - 120$),
using exposure times that should produce continuum signal-to-noise 
(S/N) ratios of
around 100 in the high resolution spectra, and 200-250 in the
low resolution spectra.  The basic calibrations were automatically 
carried out in the SSC pipeline, version S9.5, and then spectra were 
extracted and coadded interactively, using the offline pipeline.  

\begin{figure}[t!]
\epsscale{1.15}
\plotone{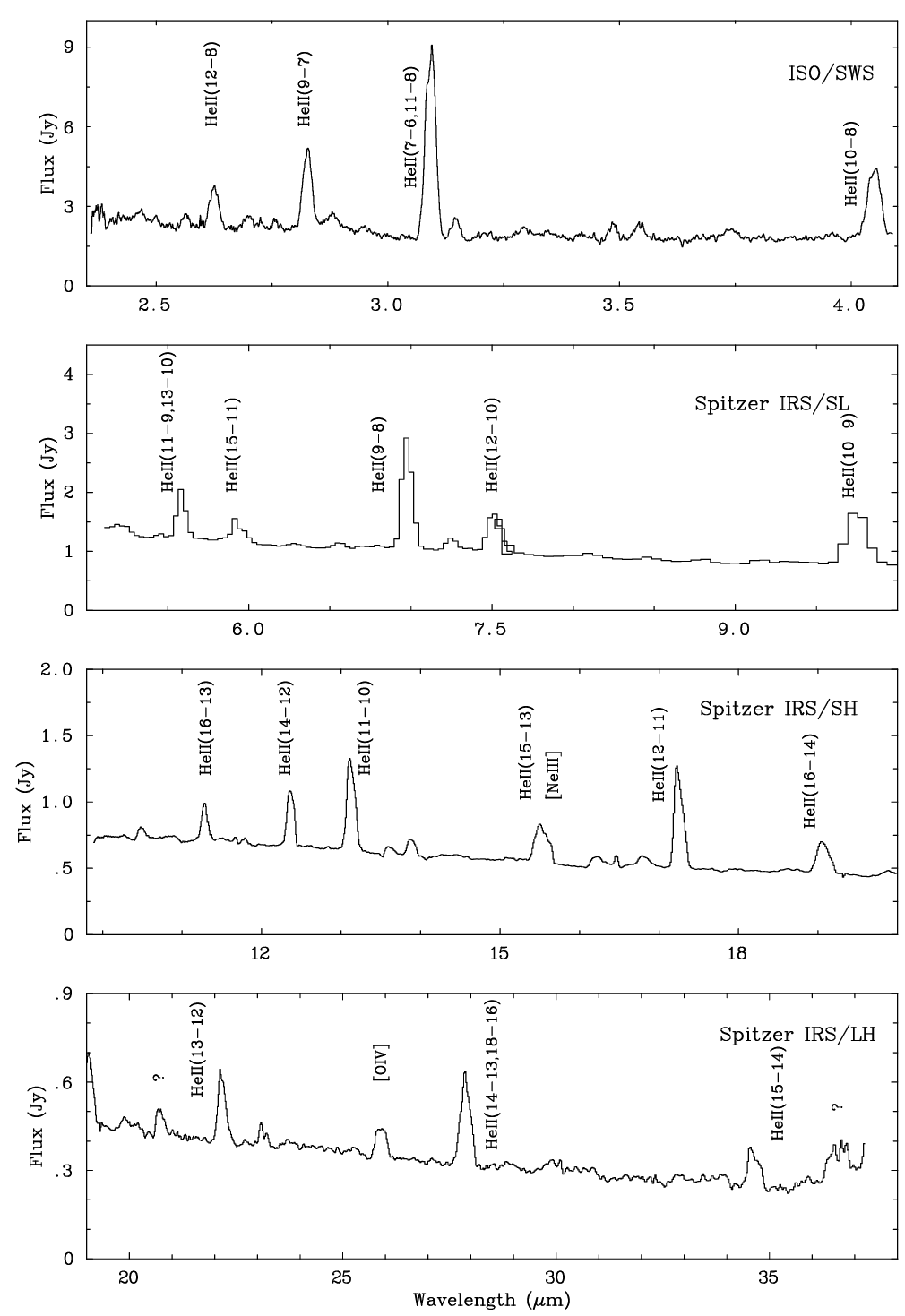}
\caption{Mid-IR spectroscopy of EZ CMa (HD~50896)
from ISO/SWS and {\em{Spitzer}}/IRS with the principal lines identified.}
\label{irs-wr6}
\end{figure}

Our analysis also makes use of archival {\em{Infrared Space Observatory}} 
({\em{ISO}}) Short Wavelength Spectrometer (SWS) spectra, optical 
spectrophotometry from Torres-Dodgen  \& Massey (1988), 
high resolution {\em International Ultraviolet Explorer (IUE)}
ultraviolet spectrophotometry from  Howarth \& Phillips (1986),
plus {\em Hopkins Ultraviolet  Telescope (HUT)} far-ultraviolet 
spectrophotometry (Schulte-Ladbeck et al. 1995). 
The unpublished SWS 2.4--4.2~$\mu$m
spectroscopy of EZ CMa was obtained from the public archive, processed to
the Auto Analysis Result.  Further reduction to a single spectrum
was carried out interactively, using SWS Interactive Analysis routines.
Continuum S/N ratios are around 50 in these data. Low resolution near-IR 
spectroscopy of HD~50896 has previously been discussed by Hillier et al. 
(1983). We  additionally use high dispersion spectroscopy of HD~50896 
obtained with the ESO 3.6m 
cassegrain echelle spectrograph (CASPEC), previously presented by
Schmutz (1997). We adopted a distance of 1.8~kpc to HD~50896 (Howarth
\& Schmutz 1995), interstellar reddening of E(B-V)=0.1 mag 
(the mean from Schmutz \& Vacca 1991, van der Hucht 2001) 
and an atomic hydrogen column density of $\log(N_{HI})=$20.7
cm$^{-2}$ (Howarth \& Phillips 1986).

The {\em{Spitzer}} IRS and {\em{ISO}} SWS spectroscopy of EZ~CMa are 
presented in Fig.~1 together with identifications of the principal
wind features. The majority of mid-IR spectral features are due to wind
lines of He\,{\sc ii}, primarily the leading member in each series
from He\,{\sc ii} (7-6) at 3.09$\mu$m to (15--14) at 34.6$\mu$m. In addition
to these, we observe fine-structure lines of [O\,{\sc iv}] 25.9$\mu$m and
[Ne\,{\sc iii}] 15.5$\mu$m, 
the latter blended with He\,{\sc ii} 15.47$\mu$m (15-13).
There is no evidence for [S\,{\sc iv}]  10.5$\mu$m
in HD~50896. The observed feature is well reproduced by a weak blend 
of He\,{\sc ii} 10.46$\mu$m (21-15) and 10.50$\mu$m (24-16) in our
synthetic spectrum.


\section{Spectroscopic analysis of HD~50896}\label{analysis}

\subsection{Technique}

To date, the majority of spectroscopic analyses of WN stars have been
carried out using non-LTE models that did 
not account for metal line blanketing or clumping 
(e.g. Crowther et al. 1995, Hamann et al. 1995). In only a few cases
have clumped, 
metal line blanketed models been applied to individual stars (e.g. 
Schmutz 1997, Morris et al. 2000,
Dessart et al. 2000,  Herald et al. 2001). Such effects need to be taken into 
account in order
to determine the fundamental parameters of WR stars (Crowther 2003).

For the present study of HD~50896, we employ
CMFGEN (Hillier \& Miller 1998), which solves the 
transfer equation in  the co-moving frame subject to statistical and 
radiative equilibrium,  
assuming an expanding, spherically-symmetric, homogeneous and static 
atmosphere, allowing for line blanketing and clumping.
The stellar radius ($R_{\ast}$) is defined
as the inner boundary of the model atmosphere and is located at 
Rosseland optical depth of $\sim$20 with the stellar temperature ($T_{\ast}$)
defined by the usual Stefan-Boltzmann relation. Consequently, the
stellar radius is much smaller than the radius of $\tau$=1, such that
the temperature depends directly on the assumed velocity law.


Our approach follows previous studies (e.g. Crowther et al. 1995), such that
diagnostic optical lines of 
He\,{\sc i} ($\lambda$5876), He\,{\sc ii} ($\lambda$4686) plus the local
continuum level allow a determination of
the stellar temperature, mass-loss rate and luminosity. We were unable to 
employ solely mid-IR diagnostics since the IRS spectrum of HD~50896 is
dominated by lines of He\,{\sc ii}, with He\,{\sc i} weak or blended.

Our final model atom contains H, He, C, N, O, Ne, Mg, Si, P, S, Cl, Ar, 
Ca, Cr, Mn, Fe and Ni. 
In total, 1186 super levels, 4462
full levels and 45,485 non-LTE transitions are simultaneously considered.
We assume hydrogen is absent, such that helium makes up 98\% of the atmosphere.
CNO abundances are varied until an optimal fit is achieved. 
In contrast with nitrogen and carbon, the abundance of oxygen in WN 
stars has proven to be rather challenging (Hillier 1988). The reason is 
that oxygen has a  much more complicated ionization stratification in 
their inner winds,  plus diagnostic lines are located in rather 
observationally inaccessible regions, typically 2800--3400\AA. Consequently
we set oxygen (and neon) abundances from the fine-structure analysis.
Other elements are fixed at Solar values (Grevesse \& Sauval 1998; 
Asplund et al. 2004). 

We adopt a standard form of the velocity law,
$v(r) = v_\infty ( 1 - R_*/r)^\beta$, where $\beta$=1. In contrast,
Schmutz (1997) solved the hydrodynamical equation for outer wind and
obtained $\beta\sim3$ with reference to the core radius. A terminal wind
velocity, $v_\infty$, of 1860 km\,s$^{-1}$ is obtained from the mid-IR
fine structure line of [O\,{\sc iv}] 25.9$\mu$m. For comparison, Prinja
et al. (1990) obtained 1720 km\,s$^{-1}$ from UV observations of the
P Cygni C\,{\sc iv} 1550\AA\ line, whilst Schmutz (1997) derived 
2060 km\,s$^{-1}$ from a fit to He\,{\sc i} 1.083$\mu$m.

The mass-loss rate is actually derived as the ratio $\dot{M}/\sqrt{f}$,
where $f$ is the volume filling factor that can be constrained by
fits to the electron scattering wings of the helium line profiles 
(following Hillier 1991). We conclude $f\sim$0.1, and can definitely 
exclude homogeneous mass-loss in HD~50896. In addition, the mid-IR 
continuum slope also reacts to different filling-factors.





\begin{table}[ht!]
\caption[]{Comparison of stellar parameters for EZ~CMa (HD~50896, WN4b)
derived here (labeled `Model') with those determined previously
by Schmutz (1997, S97), allowing for the clumped nature of the wind
in each case with a volume filling factor of $f$.}
\label{table1}
\begin{center}
\begin{tabular}{l@{\hspace{2mm}}c@{\hspace{2mm}}
l@{\hspace{2mm}}l@{\hspace{2mm}}c@{\hspace{2mm}}
l@{\hspace{2mm}}r@{\hspace{2mm}}
c@{\hspace{2mm}}
c}
\hline\noalign{\smallskip}
Analysis& $T_{\ast}$& $R_{\ast}$ & $\log L_{\ast}$&
log ({$\dot{M}\over\sqrt{f}$}) & $f$ & $\beta$ & $v_{\infty}$ & $M_{v}$\\
        & kK       & R$_{\odot}$& L$_{\odot}$    &
M$_{\odot}$ yr$^{-1}$ &  &  & km s$^{-1}$&  mag  \\
\hline\noalign{\smallskip}
Model&85&2.9& 5.58 & $-$4.0 & 0.10 &1 & 1860 & $-$4.6\\
S97  &84&3.5& 5.74 & $-$3.9 & 0.06 & $\sim$3& 2060 & $-$4.6\\
\noalign{\smallskip}\hline
\end{tabular}
\end{center}
\end{table}

\subsection{Spectroscopic Results}\label{results}

\begin{figure*}[t!]
\epsscale{0.9}
\plotone{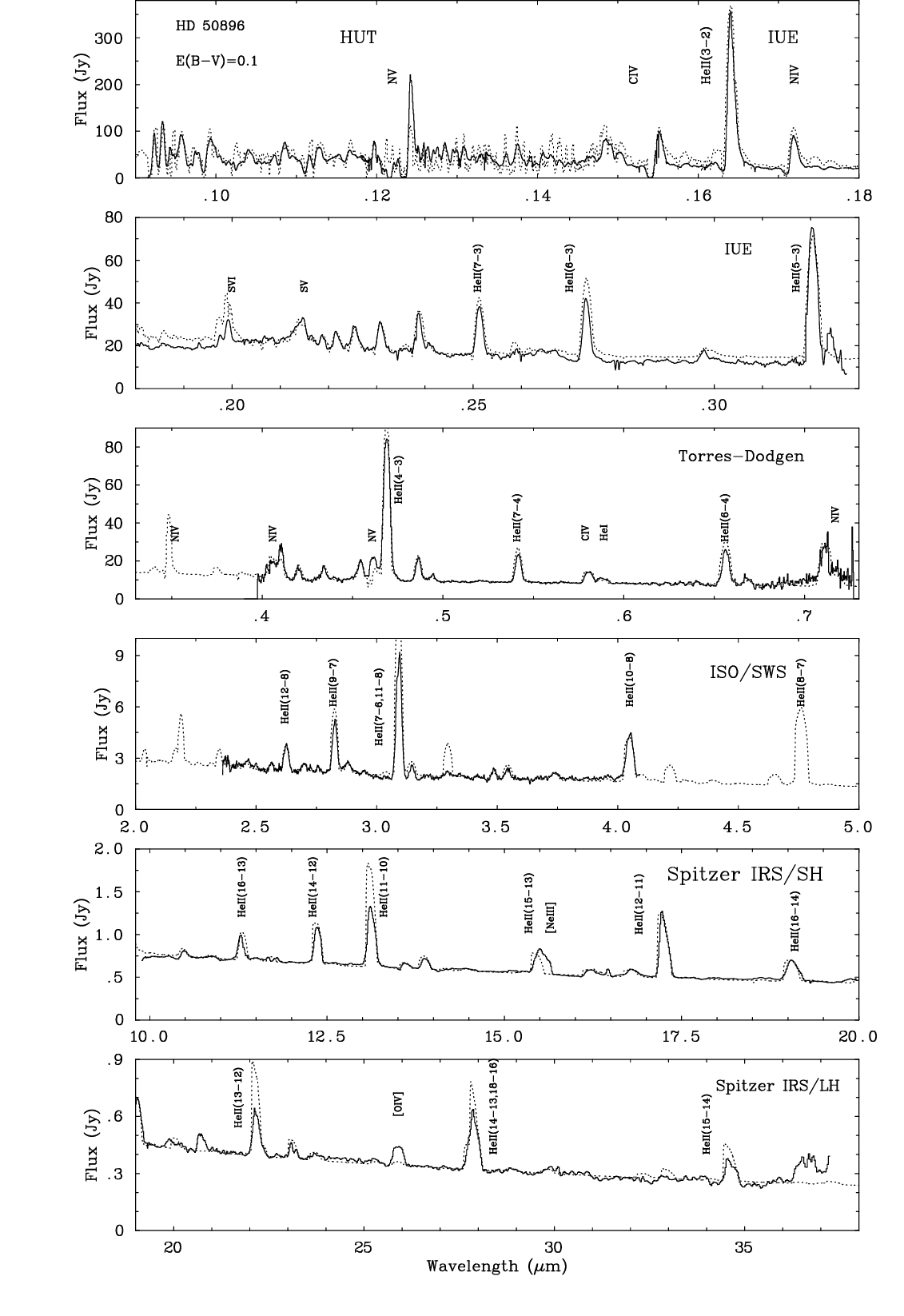}
\caption{
Comparison bwtween de-reddened far-UV (HUT), UV 
(IUE), optical (Torres-Dodgen \& Massey 1988), 
near-IR ({\em ISO}) and mid-IR ({\em{Spitzer}}) spectrophotometry of 
EZ~CMa
with our synthetic model (dotted), corrected for interstellar atomic 
hydrogen
with log(N(HI)=20.7 cm$^{-2}$ (Howarth \& Phillips 1987).
Note that the predictions from our non-LTE analysis are included here,
such that specifically the fine-structure lines are not synthesised.}
\label{wr6_fit}
\end{figure*}

We compare our synthetic model with far-UV, near-UV, optical,
near-IR and mid-IR spectrophotometry (dereddened by E(B-V)=0.10 mag)
 in Fig.~\ref{wr6_fit}. Overall the agreement between the spectral
features and continuum are excellent from 0.09--37$\mu$m, with few
major exceptions. For helium, all lines are reproduced better than 20\%, with
the exception of a few mid-IR lines, namely He\,{\sc ii} 3.091$\mu$m
(7-6), 13.12$\mu$m (11-10) and 22.17$\mu$m (13-12). Since
fine-structure lines are not included in our synthetic spectrum, 
it is apparent that
[Ne\,{\sc iii}] is a non-negligible contributor to the 15.5$\mu$m feature,
the 25.9$\mu$m line is dominated by [O\,{\sc iv}] and  there is no obvious
identification for the lines at 20.7$\mu$m or 36.6$\mu$m (the latter
is not [Ne\,{\sc iii}] 36.01$\mu$m).

The stellar parameters derived for the WN4b star are presented in 
Table~\ref{table1}. We estimate elemental abundances of N/He=2$\times 10^{-3}$
and C/He=5$\times 10^{-5}$ by number. 
Hillier (1988) previously estimated N/He$\leq4\times 10^{-3}$, N/C$\sim$14
and O/N$\leq$3 for EZ~CMa, the only previous study to attempt CNO
abundance determinations. 
%

%
%

The only  recent study of EZ~CMa allowing for blanketing
and wind clumping was by Schmutz (1997), whose results we also include in
Table~\ref{table1}. Schmutz (1997) approached the incorporation
of line blanketing in a different manner from the present approach. A Monte
Carlo sampling technique was adopted, allowing for the effect of a much more
thorough line opacity at the expense of full consistency in the radiative
transfer problem, 
via the use of approximate ionization and excitation for metal species.
In addition, Schmutz adopted a 
He\,{\sc ii} Ly$\alpha$ $\lambda$303 photon loss mechanism with a particular
parameter. Such a mechanism is intrinsic to the present analysis without the
requirement of a parameterization, providing spectral lines adjacent to 
He\,{\sc ii} $\lambda$303.78 are accounted for in the calculation. Consequently
the following metal lines within $\sim$100 
km\,s$^{-1}$ of Ly$\alpha$ are included --
O\,{\sc iii} $\lambda$303.70, 303.80, Fe\,{\sc vi} $\lambda$303.70, 303.80,
Ni\,{\sc vi} $\lambda$303.71, Mn\,{\sc vi} $\lambda$303.72, Ca\,{\sc v} 
$\lambda$303.74, Cr\,{\sc v} $\lambda$303.82, Cr\,{\sc vi} 
$\lambda$303.84 and Fe\,{\sc v}  $\lambda$303.91.

Agreement between the two approaches is reasonable, given the differences
in approach and choice of diagnostics. In both solutions, stellar luminosities 
(and hence bolometric corrections) are significantly higher  than previous 
non-LTE studies (e.g. Hamann et al. 1995), although Schmutz (1997) obtained
a rather higher bolometric luminosity, due to a combination of his photon
loss mechanism and more complete line opacity, 
whilst clumping corrected  
mass-loss rates are much lower.  Consequently, the derived momentum ratio 
is found to be  $\dot{M} v_{\infty}/(L/c)$ = 7, versus 30--70 in 
earlier studies. The agreement in mass-loss rate is consistent with the
claimed precision of 0.1 dex achieved by Schmutz (1997), 
of relevance to the determination of elemental abundances to follow.

\subsection{Elemental Abundances: Neon and Oxygen}\label{ionabun} 

\begin{figure}[t!]
\epsscale{1.1}
\plotone{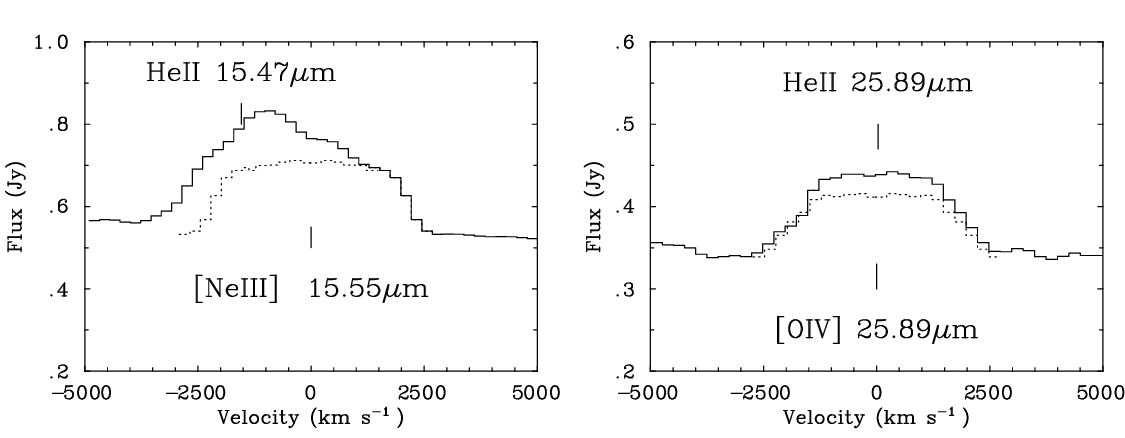}
\caption{
Mid-IR fine-structure lines comprising [Ne\,{\sc iii}] 15.55$\mu$m
and [O\,{\sc iv}] 25.9$\mu$m observed in the {\em Spitzer} IRS 
observations of EZ~CMa, together with the predicted fractional
coontribution from the fine-structure lines (dotted) themselves (see text).}
\label{wr6_fs}
\end{figure}

We present the fine-structure lines of [Ne\,{\sc iii}] 15.5$\mu$m
and [OIV] $\lambda$25.9$\mu$m in Fig.~\ref{wr6_fs}, including the
predicted fractional contributions from [Ne\,{\sc iii}] and [O\,{\sc iv}]
(dotted lines) obtained from, respectively, the red line profile of 
[Ne\,{\sc iii}]
which does not overlap with He\,{\sc ii} (15--13),
and the non-LTE predicted He\,{\sc ii} (23--19) strength.
The observed 
line fluxes of each blend is presented in  Table~\ref{finestructure}.
Sources of atomic data  are given in Dessart et al. 
(2000) for [Ne\,{\sc iii}], and Hayes \& Nussbaumer (1983) and Blum \&
Pradhan (1992) for [O\,{\sc iv}]. We have determined
elemental abundances for these ions using the  numerical 
techniques introduced by Barlow et al. (1988), and
adapted to account for a clumped wind by Dessart et al. (2000) and
Morris et al. (2000).
With regard to clumping, we admit that the 
volume filling factor in the outer wind may be considerably different
than that derived from optical and near-IR recombination lines (Runacres
\& Owocki 2002).

The numerical expression for the ion number 
fraction, $\gamma_i$, in a clumped medium (with volume filling
factor $f$) is (in cgs units)
\begin{equation}
 \gamma_{i}=\frac{(4 \pi \mu m_{H} v_{\infty})^{1.5}}{\ln(10)f^{0.25}}
\left(\frac{\sqrt{f}}{\mdot}\right)^{1.5}
\frac{1}{F_{u}(T)}\frac{2D^{2}I_{ul}}{\sqrt{\gamma_{e}}A_{ul}h\nu_{ul}} 
\end{equation}
where  $D$ is the stellar distance, $I_{ul}$ is the line flux of the 
transition with energy $h\nu_{ul}$ between upper level $u$ and lower 
level $l$, with transition probability $A_{ul}$.
$\gamma_e$ (=1.008) and $T_e$ (=14000K) are the electron number density
and temperature in the line-forming region, and the mean molecular weight
is $\mu$ (=4.04) for EZ~CMa, with $m_H$ the mass of the hydrogen atom. The
integral part, $F_u(T)$, is
\begin{equation}
F_{u}(T) = \int_{0}^{\infty}\frac{f_{u}(N_{e},T)}{\sqrt{N_{e}}}d 
\log(N_{e}).
\end{equation}
where $f_u$ is the fractional population of the upper level.

Following this technique, we derive Ne$^{2+}$/He = 1.2$\times 
10^{-4}$ and O$^{3+}$/He = 4.2$\times 10^{-5}$ by number, {\it assuming}
a fractional contribution of 67\% and 85\% to the observed
15.5$\mu$m and 25.9$\mu$m lines respectively.
If we  were to neglect the predicted He\,{\sc ii} 
contribution to the observed spectral features, we would
obtain Ne$^{2+}$/He = 1.8$\times 10^{-4}$ and O$^{3+}$/He = 
4.8$\times 10^{-5}$ from this method. 
Applying the same technique
based on the Schmutz (1997) spectroscopic results, we would derive
slightly lower abundances of
Ne$^{2+}$/He = 1.1--1.4$\times 10^{-4}$ and O$^{3+}$/He = 
3.5--4.1$\times 10^{-5}$.




Of course, one needs to account for potential contributions from unseen
ionization stages. Our model atmosphere identifies the following dominant
ions at the outer model radius of $\sim$200$R_{\ast}$, where
$\log (n_{e}/$cm$^{3})\sim$9, namely
He$^{+}$, C$^{3+}$, N$^{3+}$, O$^{3+}$, Ne$^{2+}$, Mg$^{2+}$, Si$^{4+}$, 
P$^{4+}$, S$^{4+}$, Cl$^{4+}$, Ar$^{4.5+}$, Ca$^{4+}$, Cr$^{4+}$, 
Mn$^{5+}$, Fe$^{4.5+}$ and Ni$^{5+}$. 
The dominant ionization state in WN stars  at large radii is predicted to
be higher than for WC stars, despite similar stellar temperatures, 
since the effects of metal coolants is far less (Hillier 1988, 1989).
Are these predictions
consistent for the physical conditions at even lower densities of 
$\log (n_{e}/$cm$^{3})\sim$5 where the fine-structure lines form? 

The absence of [S\,{\sc iv}] 10.5$\mu$m is consistent with a dominant
ionization stage of S$^{4+}$ for EZ~CMa, whilst
[Ne\,{\sc ii}] 12.8$\mu$m line is absent,
so Ne$^{+} \ll$ Ne$^{2+}$, in agreement with expectation. 
The contribution of Ne$^{3+}$ is uncertain, although it has a rather
high ionization potential (IP = 97eV), such that we
assume  Ne/He$\approx$ Ne$^{2+}$/He = 1.2--1.8
$\times 10^{-4}$, depending on the contribution by He\,{\sc ii}. 
This is in reasonable agreement with the revised Solar 
Ne abundance (Asplund et al. 2004) of Ne/He=2$\times 10^{-4}$, 
adjusted for the H-depleted atmosphere of an early WN star. 

\begin{table}[t!]
\caption{Observed mid-IR fine structure line intensities (units of
10$^{-12}$ erg~cm$^{-2}$~s$^{-1}$) in EZ~CMa (uncorrected for potential
blends with He\,{\sc ii} lines),
and adopted atomic parameters, including
statistical weights of the upper and lower levels, $\omega_u$ and 
$\omega_l$,
transition probability, $A_{ul}$, and collision strength $\Omega_{ul}$ at
$T_e$=14000K.}
\label{finestructure}
\begin{center}
\begin{tabular}{lccccccc}
\hline\noalign{\smallskip}
Ion & Transition & $\lambda$ & $\omega_u$ & $\omega_l$ & $A_{ul}$ & 
$\Omega_{ul}$ & $I_{ul}$(blend) \\
\hline\noalign{\smallskip}
{}[Ne\,{\sc iii}] & $^3$P$^{\circ}_{1}$--$^3$P$^{\circ}_{2}$&
15.55 & 3 & 5 & 5.99$\times$10$^{-3}$&1.65& 7.45$\times 10^{-13}$ \\
{}[O\,{\sc iv}]   & $^2$P$^{\circ}_{1/2}$--$^{2}$P$^{\circ}_{3/2}$ &
25.89 & 4 & 2 & 5.19$\times$10$^{-4}$&2.52& 1.49$\times 10^{-13}$ \\
\noalign{\smallskip}\hline
\end{tabular}
\end{center}
\end{table}
{}For oxygen, we expect that O$^{3+}$ is the dominant ionization
stage, but are unable to verify the absence of fine-structure O$^{2+}$
lines, which lie at 51.8 and 88.2$\mu$m, longward of the IRS passband.
Nevertheless, from the consistency with other relevant ions of Ne and S,
we adopt O/He $\simeq$ O$^{3+}$/He = 4.2--4.8$\times$10$^{-5}$. 
{}From our nitrogen
abundance derived above, we find N/O=40--48 by number. What is the
expected oxygen abundance in a H-free WN star?
{}From the spectroscopic analysis above, we derive N/C=40 by number. This may be
compared with recent predictions for (initially) rotating massive stars
(Meynet \& Maeder 2003). At the end of H-burning, the 60--120$M_{\odot}$
tracks for initial equatorial rotation velocities of 300 km\,s$^{-1}$ predict 
N/C=40--45 by number, in perfect agreement. 
At this stage, N/O=30--50 by number, such that the measured value lies 
in precisely the predicted range.





\section{Summary}\label{summary}

We present {\em Spitzer} IRS spectroscopy for EZ~CMa (HD~50896
WN4b), the first 
early WN star to be observed spectroscopically in the mid-IR. In addition
to numerous stellar wind lines of He\,{\sc ii}, fine structure lines of 
[Ne\,{\sc iii}] and [O\,{\sc iv}] are revealed, permitting an 
estimate of the asymptotic wind velocity, $\sim$1860 km\,s$^{-1}$, plus
elemental abundances. We carry our a spectroscopic analysis of HD~50896,
allowing for metal line blanketing and wind clumping, revealing generally
excellent agreement with the IRS spectroscopy, plus stellar parameters 
supporting the earlier study of Schmutz (1997), based on an alternative
line blanketing technique. We also derive N/He$\sim 2\times 10^{-3}$ by
number, with N/C$\sim$40.  From the derived clumped mass-loss rate, we
obtain Ne$^{2+}$/He=1.2$\times 10^{-4}$, allowing for
the contribution by neighboring He\,{\sc ii} wind lines, in comparison 
with  Ne/He=2$\times 10^{-4}$ for a severely H-depleted 
environment, adjusted for the recent downward revision in the Solar neon 
abundance by Asplund et al. (2004). An indication of the uncertainty in 
the fine-structure
derived Ne abundance may be obtained by adopting the spectroscopic 
results for EZ~CMa from Schmutz (1997), leading to a 10--20\% difference.
 In addition, we obtain O$^{3+}$/He=4.2$\times 10^{-5}$, i.e. N/O$\sim$48, 
 in excellent agreement with evolutionary predictions for massive stars
at the end of core H-burning. 
%

%

\acknowledgements
We are of course 
grateful to John Hillier for the use of his non-LTE atmospheric
code and to the referee Werner Schmutz for his constructive comments.   







\begin{thebibliography}{}
\bibitem[]{703} Asplund M., Grevesse N., Sauval A.J., Allende Prieto C., 
Kiselman D., 2004, A\&A 417, 751
\bibitem[]{669} Barlow M.J., Roche P.F,  Aitken D.A., 1988, MNRAS 232, 821 
\bibitem[]{670} Blum R.D., Pradhan A.K., 1992, ApJS 80, 425
\bibitem[]{671} Crowther P.A., 2003, in Proc IAU Symp 212, A Massive Star 
Odyssey, from Main Sequence to Supernova, eds. K.A. van der Hucht et al., 
(ASP: San Francisco) p.47
\bibitem[]{674} Crowther, P.A., Hillier, D.J., Smith, L.J. 1995, A\&A, 293, 407
\bibitem[]{675} Dessart L., Crowther P.A., Hillier, D.J. et al., 2000, 
MNRAS 315, 407
\bibitem[]{677} Grevesse N., Sauval A.J., 1998, Space Sci. Rev. 85, 161
\bibitem[]{679} Hayes M.A., Nussbaumer H., 1983, A\&A 124, 279
\bibitem[]{680} Hamann W-R., Koesterke L.,  Wessolowski U., 1995, A\&A 299, 151
\bibitem[]{681} Herald J., Hillier D.J., Schulte-Ladbeck R.E., 2001, ApJ 548, 932
\bibitem[]{682} Hillier D.J., 1988, ApJ 327, 822
\bibitem[]{683}  Hillier D.J., 1989, ApJ 347, 392
\bibitem[]{684} Hillier D.J., 1991, A\&A 247, 455
\bibitem[]{685} Hillier, D.J., Miller, D.L., 1998, ApJ, 496, 407
\bibitem[]{686} Hillier D..J., Jones T.J., Hyland A.R., 1983, ApJ 271, 221
\bibitem[]{686a} Houck, J.R., et al. 2004, \apjs, this volume.
\bibitem[]{687} Howarth I.D., Phillips A.P., 1986, MNRAS 222, 809
\bibitem[]{688} Howarth I.D., Schmutz W.,  1995, A\&A 294, 529 
\bibitem[]{689} van der Hucht K.A., 2001, New Astron. 35, 145
\bibitem[]{690} Meynet, G., Maeder, A., 2003, A\&A, 404, 975
\bibitem[]{691} Morris P.W., van der Hucht, K.A., et al. 2000, A\&A 353, 624
\bibitem[]{692} Runacres M.C., Owocki S.P., 2002, A\&A 381, 1015
\bibitem[]{693} Schmutz W., 1997, A\&A 321, 268
\bibitem[]{694} Schmutz W., Vacca W.D., 1991, A\&A 248, 678
\bibitem[]{695} Schulte-Ladbeck R.E., Hillier D.J., Herald J.E., 1995, ApJ 454, 
L51
\bibitem[]{697} Torres-Dodgen A.V., Massey P., 1988, AJ 96, 1076
\end{thebibliography}
\end{document}